\documentclass[12pt,onecolumn]{IEEEtran}
 \usepackage[T1]{fontenc}
\usepackage{ae,aecompl}

\usepackage{amsmath}
\usepackage{theorem}
\usepackage{psfig}
\usepackage{epsfig}
\usepackage{amsfonts}
\usepackage{amssymb}
\usepackage{cite}

\renewcommand{\baselinestretch}{1}

\renewcommand{\baselinestretch}{1.45}

\newtheorem{theorem}{Theorem}
\newtheorem{lemma}{Lemma}
\newtheorem{corollary}{Corollary}

\newcommand{\comment}[1]{}

\begin{document}
\thispagestyle{empty} \vskip 1cm
\begin{figure}[htb]
\centerline{\psfig{figure=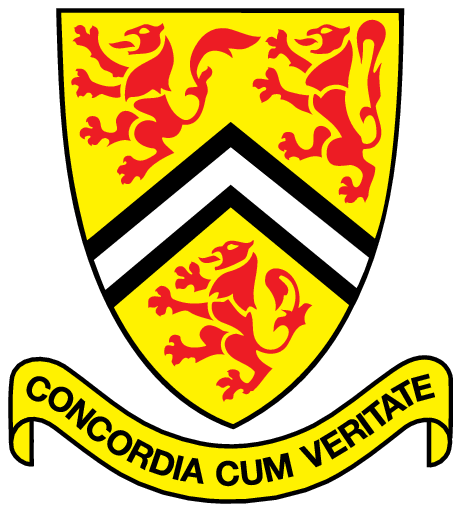} }
\end{figure}
\vspace{1cm}

\begin{center}
\vskip 0.6cm
{\large \bf Characterization of Rate Region in Interference \\ Channels with Constrained Power} \\

\vskip 0.8cm

Hajar Mahdavi-Doost, Masoud Ebrahimi, and Amir K. Khandani

\vskip 0.5cm

{\small Coding \& Signal Transmission Laboratory\\
Department of Electrical \& Computer Engineering\\
University of Waterloo\\
Waterloo, Ontario, Canada, N2L 3G1\\

Technical Report UW-E\&CE\ 2006-19\\} September. 19, 2006

\end{center}

\pagenumbering{arabic} \setcounter{page}{0}

\title{\bf Characterization of Rate Region in Interference Channels with Constrained Power
\footnote{}
\thanks{This work is financially supported by
Nortel Networks, National Sciences and Engineering Research Council
of Canada (NSERC), and Ontario Centres of Excellence (OCE).}}
\author{Hajar Mahdavi-Doost, Masoud Ebrahimi, and Amir K. Khandani \\
\small Coding \& Signal Transmission Laboratory(www.cst.uwaterloo.ca)\\
Dept. of Elec. and Comp. Eng., University of Waterloo\\ Waterloo, ON, Canada, N2L 3G1 \\
e-mail: \{hajar, masoud, khandani\}@cst.uwaterloo.ca}
\maketitle
\begin{abstract}
In this paper, an $n$-user Gaussian
interference channel, where the power of the transmitters are
subject to some upper-bounds is studied. We obtain a closed-form expression
for the rate region of such a channel based on the
Perron-Frobenius theorem. While the boundary of the rate region
for the case of unconstrained power is a well-established result,
this is the first result for the case of constrained power. We
extend this result to the time-varying channels and obtain a
closed-form solution for the rate region of such channels.
\end{abstract}

\section{Introduction}
%
%In an interference channel, a number of source nodes transmit data
%to their designated destination nodes through a shared wireless
%channel\cite{Cover}. Many systems like cellular networks, sensor
%networks, and ad-hoc networks fall in this category. Since the
%capacity of such a channel has not been characterized yet, it has
%been investigated under some simplifying conditions. A typical
%assumption  is to treat the interference as a Gaussian noise. In
%this case, the Shannon capacity would be proportional to the
%logarithm of signal-to-interference-plus-noise ratio (SINR) plus
%one. We will follow this assumption throughout this paper.

Channel sharing is known as an efficient scheme to increase the
spectral efficiency of the wireless systems. While such a scheme
increases the capacity and the coverage area of systems, it
suffers from the interference among the concurrent links (co-channel interference). Consequently, the
signal-to-interference-plus-noise-ratio (SINR) of the links are
upper-bounded, even if there is no constraint on the transmit
powers.

There have been some efforts to evaluate the maximum achievable
SINR in the interference channels. In~\cite{Aein_balancing}, the maximum achievable SINR of a
system with no constraint on the power is expressed in terms of the Perron-Frobenius
eigenvalue of a non-negative matrix and this result is utilized to develop an SINR-balancing
scheme for satellite networks. This formulation for the maximum
achievable SINR is deployed in many other wireless communication
applications such as \cite{Alavi_balancing,
Zander_performance,Zander_distributed,Hanly_powercontrol}
afterwards.

Recently, the rate region of interference channels and its
properties has been investigated in the literature. In
\cite{Catrein_duality}, it is shown that the capacity region when
the power is unbounded is convex. The capacity region in
\cite{Catrein_duality} is defined as the set of feasible
processing gains while for a constant bandwidth, the processing gain is
inversely proportional to the rate. In \cite{Imhof_power}, some
topological properties of the capacity region (with the
aforementioned definition) of CDMA systems are investigated for
the cases when there are constraints on the power of individual
users and when there is no constraint on the power. The authors in
\cite{Imhof_power} show that the boundary of the capacity region
with one user's power fixed and the rest unbounded is a shift of
the boundary of some capacity region with modified parameters, but
unlimited power. However, this result is not in a closed form and
can not be extended for the other forms of power constraints.
%Using supporting hyperplanes and topological arguments, the
%convexity of the capacity region with limited power follows.

It is shown that the feasible SINR region is not convex, in
general \cite{Boche_convexity,Boche_totalpower,Cruz_optimal}. In \cite{Sung_region}, it is shown that in
the case of unlimited power, the feasible SINR region is
log-convex. The authors in \cite{Catrein_duality} also consider a
CDMA system without power constraints, and show that the
feasible inverse-SINR region is a convex set. In
\cite{Boche_convexity}, it is proved that the feasible quality of service (QoS) region
is a convex set, if the SINR is a log-convex function of the
corresponding QoS parameter. Reference \cite{Boche_infeasible}
shows that under a total power constraint, the infeasible SINR
region is not convex.

%In \cite{Boche_totalpower}, the convexity results of
%\cite{Catrein_duality} is extended to systems with the total power
%constraint.

In this paper, we extend the result on the maximum achievable SINR
in \cite{Aein_balancing} to the systems with certain constraints
on the power of transmitters. This  result which is based on
Perron-Frobenius theorem, yields a closed-form solution for the
rate region of the systems with constrains on the power. The
extendable structure of constraints enables us to use the proposed
derivation for the maximum achievable SINR in many practical
systems. This result is extended to a time-varying system, where
the channel gain is selected from a limited-cardinality set, and
the average power of users are subject to some upper-bounds.

{\em Notation:} All boldface letters indicate column vectors (lower
case) or matrices (upper case). $x_{ij}$ and $\mathbf{x}_i$
represent the entry $(i,j)$ and column $i$ of the matrix
$\mathbf{X}$, respectively. A matrix $\mathbf{X}_{n\times m}$ is
called \textit{non-negative} if $x_{ij} \geq 0,\ \forall i,j$, and
denoted by $\mathbf{X} \geq \mathbf{0}$. Also, we have
$$\mathbf{X}\geq \mathbf{Y} \Longleftrightarrow
\mathbf{X}-\mathbf{Y} \geq \mathbf{0},$$ where $\mathbf{X},
\mathbf{Y}$ and $\mathbf{0}$ are non-negative matrices of compatible
dimensions \cite{Seneta}. $\det(\mathbf{X})$,
$\mathrm{Tr}(\mathbf{X})$, $\mathbf{X}'$, and $|\mathbf{X}|$ denote
the determinant, the trace, the transpose, and the norm of the
matrix $\mathbf{X}$, respectively. $\mathbf{I}$ is an identity
matrix with compatible size. $\otimes$ represents the Kronecker
product operator. $\textrm{diag}(\mathbf{x})$ is a diagonal matrix
whose main diagonal is $\mathbf{x}$. We define the reciprocal of
polynomial $\mathrm{q}(x)$ of degree $m$ as
$x^m\mathrm{q}(\frac{1}{x})$.
$\psi(\mathbf{X},\mathbf{y},\mathcal{S})$ is a matrix defined as a
function of three parameters, which are respectively a matrix, a
vector and a set of indices,
\begin{equation}\nonumber
%\label{Xi_definition}
\psi(\mathbf{X},\mathbf{y},\mathcal{S})=\mathbf{Z}=[\mathbf{z}_j],
\quad \mathbf{z}_j= \left\{
\begin{array}{ll}
\mathbf{x}_j+\mathbf{y} & j\in \mathcal{S}
\\
 \mathbf{x}_j & \mathrm{otherwise}
\end{array} \right.
\end{equation}

%======================================================================
\section{Problem Formulation}
\begin{figure}[bmpt]
\centerline{\psfig{figure=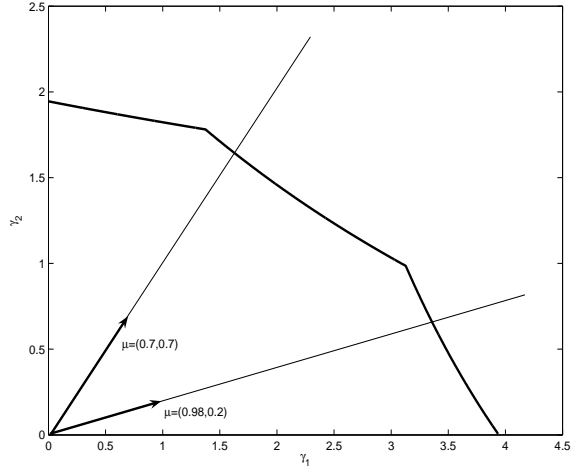,width=3 in,height=2.5 in}}
\caption[The boundary of SINR Region for an interference channel with $2$
users]{\small{The boundary of SINR Region for an interference channel with $2$
users } } \label{sr_min}
\end{figure}

%======================================================================
An interference channel, including $n$ links (users), is
represented by the gain matrix $\mathbf{G}=[g_{ij}]_{n \times n}$
where $g_{ij}$ is the attenuation of the power from
transmitter $j$ to receiver $i$. This attenuation can be the
result of fading, shadowing, or the processing gain of the CDMA
system. A white Gaussian noise with zero mean and variance
$\sigma_i^2$ is added to each signal at the receiver $i$ terminal.
In many applications, the  QoS of the system is measured by an
increasing function of SINR. In an interference channel, SINR of each user, denoted by $\gamma_i$, is
\begin{eqnarray}
\nonumber
%\label{gamma_i}
&& \gamma_i=\dfrac{g_{ii}p_{i}}{\sigma_i ^2+\displaystyle\sum_{\substack{j=1 \\
j\neq i}}^{n}{g_{ij}p_j} }  , \quad  \forall i \in \{1, \ldots ,
n\},
\end{eqnarray}
where $p_i$ is the power of transmitter $i$. In addition, in
practice, the power vector $\mathbf{p}$ is subject to a set of
constraints. The main goal is to find the maximum SINR which can be obtained by all users in
the presence of such constraints. To this end, we solve the
following optimization problem
\begin{align}
 \label{main_problem} &\max \gamma \\
\label{main_constraint}  \mathrm{s.t.} \quad &\gamma_i \geq \mu_i \gamma \\
\label{general_constraint1}
&\mathbf{p}\geq \mathbf{0} \\
\label{general_constraint2} &\displaystyle\sum_{i\in \Omega}{p_i} \leq \overline{p}_{\Omega},
\end{align}
where $\Omega \subseteq \{1, \ldots ,n \}$ with $k$ elements and
$\boldsymbol{\mu}$ is a given vector with $\mu_i\geq 0$ and $|
\boldsymbol{\mu} |=1$. As we will see, the solution can be easily extended for the case of multiple power constraints of the form $\displaystyle\sum_{i\in \Omega}{p_i} \leq \overline{p}_{\Omega}$ for different $\Omega \subseteq \{1, \ldots n\}$. $\boldsymbol{\mu}$ provides the flexibility
of satisfying different rate services for different users.
According to Fig. \ref{sr_min}, the solution of
\eqref{main_problem} yields the maximum achievable SINR in the
direction of vector $\boldsymbol{\mu}$. Although the numerical
solution of this problem is already obtained through geometric
programming \cite{Geometric_communication},
\cite{Geometric_example}, we propose a different approach which leads
to a closed-form result.

%\section{Maximum Achievable SINR}

By defining the normalized gain matrix $\mathbf{A}$ as
\begin{equation}
%\label{A_definition}
\nonumber \mathbf{A}=[a_{ij}]_{n \times n}, \quad a_{ij}=\left \{
\begin{array}{ll}\dfrac{g_{ij}}{g_{ii}} &i\neq j \\ 0 &i=j \end{array} \right.
\end{equation}
the constraint \eqref{main_constraint} is rewritten as
\begin{equation}\label{main_constraint1}
\dfrac{p_{i}}{\eta_i+\displaystyle\sum_{j=1}^{n}{\mu_ia_{ij}p_j} }
\geq  \gamma , \quad \forall i \in \{1, \ldots , n\},
\end{equation}
where \begin{equation}\label{e_definition} \eta_i=\dfrac{\mu_i
\sigma_i ^2}{g_{ii}} , \quad \boldsymbol{\eta}=[\eta_i]_{n\times
1}. \end{equation} Since we are interested in maximizing the
minimum SINR, if SINR of one user is more than that of the others,
it can reduce its power to other users' advantage, and finally the
minimum SINR is improved. Therefore, equality holds in
\eqref{main_constraint1} as
\begin{equation}
%\label{linear_equation}
\nonumber
\dfrac{p_{i}}{\eta_i+\displaystyle\sum_{j=1}^{n}{\mu_ia_{ij}p_j} }
= \gamma , \quad  \forall i \in \{1, \ldots , n\}.
\end{equation}
After reformulating the problem in a
matrix form we will have
\begin{equation}\label{matrix_p}
\big(\dfrac{1}{\gamma}\mathbf{I}-\mathrm{diag}{(\boldsymbol{\mu})}\mathbf{A}\big)\mathbf{p}=\boldsymbol{\eta}.
\end{equation}
The objective is to find the maximum $\gamma$ while the system of
linear equations in \eqref{matrix_p} yields a power satisfying the
constraints on the power vector \eqref{general_constraint1},
\eqref{general_constraint2}.

When there is no constraint on the power vector (rather than
trivial constraint of $\mathbf{p}\geq\mathbf{0}$), the maximum
achievable SINR, $\gamma^*$, is characterized based on the
Perron-Frobenius theorem as
\begin{equation}\label{max_u}
\gamma^*=\dfrac{1}{\lambda^*{\big(\mathrm{diag}{(\boldsymbol{\mu})}\mathbf{A}\big)}}.
\end{equation}
where $\lambda^*$ is the Perron-Frobenius eigenvalue of the
associated matrix  \cite{Seneta}. This result was deployed in the
communication systems for
SINR-balancing ($\mu_1=\mu_2=\ldots=\mu_n$) in \cite{Aein_balancing}
for the first time.

%When the
%system is noiseless, the eigenvector corresponding to the
%PF-eigenvalue of $\mathbf{A}$ would be the power vector of the
%system which achieves $\gamma^*$. This property is utilized in
%\cite{Zander_distributed} to develop a distributed SIR-balancing
%algorithm.
We find the maximum achievable SINR, considering
certain upper-bounds on the power of transmitters in the following
sections.
%%%%%%%%%%%%%%%%%%%%%%%%%%%%%%%%%%%%%%%%%%%%%%%%%%%%%%%%%%%%%%%%%%%%%%%%%%%%%%%%
\section{SINR Region Characterization} \label{general_section}

We define $\mathbf{F}$ as
\begin{equation}\label{F_definition}
\mathbf{F}=\mathbf{I}-\gamma\mathrm{diag}{(\boldsymbol{\mu})}
\mathbf{A}.
\end{equation}
Then, the system of linear equations in \eqref{matrix_p} is
reformulated as
\begin{equation}\label{FP=eta}
\mathbf{Fp}=\gamma\boldsymbol{\eta},
\end{equation}
where $\boldsymbol{\eta}$ is defined in \eqref{e_definition}.
According to the Cramer's rule, the solution to \eqref{FP=eta} is
obtained by
\begin{equation}
%\label{pi_cramer}
\nonumber p_i=\dfrac{\det(\mathbf{H}^{(i)})}{\det(\mathbf{F})},
\end{equation}
where
\begin{equation}\label{H_definition}
\mathbf{H}^{(i)}=[\mathbf{{h}}^{(i)}_{j}]_{n \times n} , \quad
\mathbf{{h}}^{(i)}_{j}=\left \{ \begin{array}{ll}
\gamma\boldsymbol{\eta}
&j=i \\
\mathbf{{f}}_{j}  &j \neq i \end{array} \right..
\end{equation}

\noindent Defining
$
%\label{mathrmh}
\mathrm{h}^{(i)}(\gamma)=\det(\mathbf{H}^{(i)}) \quad \mathrm{and}
\quad  \mathrm{f}(\gamma)=\det(\mathbf{F}),
$
we have $$
%\label{pi_frac}
p_i=\dfrac{\mathrm{h}^{(i)}(\gamma)}{\mathrm{f}(\gamma)}.
$$ Therefore, the constraint in \eqref{general_constraint2} can be
written as
\begin{equation}
\label{general_1}
\dfrac{\displaystyle\sum_{i\in\Omega}{\mathrm{h}^{(i)}(\gamma)}}{\mathrm{f}(\gamma)}\leq
\overline{p}_{\Omega}.
\end{equation}
Defining
$$\mathrm{u}_{\Omega}(\gamma)=\overline{p}_{\Omega}\mathrm{f}(\gamma)-\displaystyle\sum_{i\in\Omega}{\mathrm{h}^{(i)}(\gamma)},$$
the inequality \eqref{general_1} is equivalent to
\begin{equation} \label{num_denom}
\dfrac{\mathrm{u}_{\Omega}(\gamma)}{\mathrm{f}(\gamma)} \geq 0.
\end{equation}
We desire to find the largest possible interval where both the
numerator and the denominator have the same sign. It can be shown that this interval is connected and adjacent to zero.
Apparently, $ \mathrm{u}_{\Omega}(0)>0, \quad \mathrm{and} \quad
\mathrm{f}(0)>0.$ Consequently, $$ \exists \quad \epsilon >0 \quad
: \quad  \mathrm{f}(\epsilon)> 0 \quad \mathrm{and} \quad
\mathrm{u}_{\Omega}(\epsilon)>0 .$$ Therefore, both the numerator
and the denominator are positive in the positive neighborhood of
zero. For satisfying \eqref{num_denom}, we have to find the
smallest positive real simple root of the numerator and the denominator, $r(\mathrm{u}_{\Omega})$ and $r(\mathrm{f})$,
and take the minimum of the two as
\begin{equation}\label{min(f,u_general)}
\hat{\gamma}=\min{\{r(\mathrm{f}),r(\mathrm{u}_{\Omega})\}} .
\end{equation}
For the sake of simplicity, without loss of generality, we assume
that $\Omega=\{1, \ldots, k\}$, $k \leq n$, i.e., the first $k$ users
are subject to the total power constraint. For the numerator we
have
\begin{align}
%\label{total_u}
\nonumber
\mathrm{u}_{\Omega}(\gamma)&=\overline{p}_{\Omega}\det{(\mathbf{F})}-\displaystyle\sum_{i=1}^{k}{\det{(\mathbf{H}^{(i)})}} \\
\label{u_last_line} &=\overline{p}_{\Omega}\big(\det{(
\mathbf{F})}
-\displaystyle\sum_{i=1}^{k}{\det{(\mathbf{\hat{H}}^{(i)})}}\big),
\end{align}
where $\mathbf{\hat{H}}^{(i)}$ is defined as $$
%\label{Hhat_definition}
\mathbf{\hat{H}}^{(i)}=[\mathbf{\hat{h}}^{(i)}_j]_{n \times n},
\quad \mathbf{\hat{h}}^{(i)}_j=\left\{ \begin{array}{ll}
\dfrac{\gamma\boldsymbol{\eta}}{\overline{p}_{\Omega}}  &j=i
\\ \mathbf{{f}}_j  &j \neq i
\end{array} \right. .
$$
\begin{lemma}\label{summation of matrices}
If square matrices $\mathbf{X}$ and $\mathbf{Y}$ differ only in
column $i$, i.e.,
$$\nonumber
\left\{
\begin{array}{l}
\mathbf{x}_{j} \neq \mathbf{y}_{j} \quad j=i\\
\mathbf{x}_{j} = \mathbf{y}_{j}\quad j \neq i
\end{array}
\right. ,$$
then
\begin{align}
\nonumber
\det\big(\mathbf{X})+\det(\mathbf{Y}\big)&=\det{\big(\psi(\mathbf{X},
\mathbf{y}_i,\{i\})\big)}\\ \nonumber
                                 &=\det{\big(\psi(\mathbf{Y},
                                 \mathbf{x}_i,\{i\})\big)}.
\end{align}
\end{lemma}
Equation \eqref{u_last_line} is rewritten as
\begin{equation}
\label{u_lastline}
\mathrm{u}_{\Omega}(\gamma)=\overline{p}_{\Omega}\big(\det{(
\mathbf{F})}-\det(\mathbf{\hat{H}}^{(1)})-\displaystyle\sum_{i=2}^{k}{\det{(\mathbf{\hat{H}}^{(i)})}}\big).
\end{equation}
Since $\mathbf{F}$ and $\mathbf{\hat{H}}^{(1)}$ are the same
except for the first column, using Lemma \ref{summation of
matrices} , we will have
\begin{align}\label{F-H}
\det{(\mathbf{F})}-\det(\mathbf{\hat{H}}^{(1)})=\det{\big(\psi(\mathbf{F},-\dfrac{\gamma\boldsymbol{\eta}}{\overline{p}_{\Omega}},\{1\})\big)}.
\end{align}
On the the other hand, using the fact that addition or substraction
of columns does not change the value of the determinant, we will
have
\begin{equation}\label{det_Hhat}
\det{(\mathbf{\hat{H}}^{(i)})}=\det\big(\psi(\mathbf{\hat{H}}^{(i)},-\mathbf{\hat{h}}^{(i)}_i,\{1,
\ldots,i-1\})\big).
\end{equation}
Then, using \eqref{F-H} and \eqref{det_Hhat} and regarding
$\mathbf{\hat{h}}^{(i)}_i=\dfrac{\gamma\boldsymbol{\eta}}{\overline{p}_{\Omega}},$
we can rewrite \eqref{u_lastline} as
\begin{align}\label{successive}
\mathrm{u}_{\Omega}(\gamma)&=\overline{p}_{\Omega}\Big(\det{\big(\psi(\mathbf{F},-\dfrac{\gamma\boldsymbol{\eta}}{\overline{p}_{\Omega}},\{1\})\big)}\\
\nonumber
&-\displaystyle\sum_{i=2}^{k}{\det(\psi(\mathbf{\hat{H}}^{(i)},-\dfrac{\gamma\boldsymbol{\eta}}{\overline{p}_{\Omega}},\{1,
\ldots,i-1\})) } \Big).
\end{align}
Since $\mathbf{F}$ and $\mathbf{\hat{H}}^{(i)}$ are the same
except for the column $i$, we can easily see that
$\psi(\mathbf{F},-\dfrac{\gamma\boldsymbol{\eta}}{\overline{p}_{\Omega}},\{1,
\ldots, i-1\})$ and
$\psi(\mathbf{\hat{H}}^{(i)},-\dfrac{\gamma\boldsymbol{\eta}}{\overline{p}_{\Omega}},\{1,
\ldots,i-1\})$ are the same except for the $i^{th}$ column. Therefore,
\begin{align}
\nonumber
&\det\big(\psi(\mathbf{F},-\dfrac{\gamma\boldsymbol{\eta}}{\overline{p}_{\Omega}},\{1,
\ldots, i-1\})\big)\\
\nonumber &-\det\big(\psi(\mathbf{\hat{H}}^{(i)},-\dfrac{\gamma\boldsymbol{\eta}}{\overline{p}_{\Omega}},\{1,\ldots,i-1\})\big)\\
\nonumber
&=\det{\big(\psi(\mathbf{F},-\dfrac{\gamma\boldsymbol{\eta}}{\overline{p}_{\Omega}},\{1,
\ldots,i\})\big)}.
\end{align}
Applying this result to \eqref{successive} successively yields the following lemma.

\begin{lemma} \label{lemma_u} $$\mathrm{u}_{\Omega}(\gamma)=\overline{p}_{\Omega}\det{\big(\psi(\mathbf{F},-\dfrac{\gamma\boldsymbol{\eta}}{\overline{p}_{\Omega}},\Omega)\big)}.
$$
\end{lemma}
We utilize the result in Lemma \ref{lemma_u} to find the smallest
positive simple root of $\mathrm{u}_{\Omega}$ using
Perron-Frobenius theorem. This theorem states some properties
about the eigenvalues of a primitive matrix. A square non-negative
matrix $\mathbf{X}$ is said to be primitive if there exists a
positive integer $k$ such that $\mathbf{X}^k>\mathbf{0}$
\cite{Seneta}.
\begin{theorem} \cite{Seneta} (The Perron-Frobenius Theorem for primitive matrices)\label{PF}
Suppose $\mathbf{X}$ is an $m \times m$ non-negative primitive
matrix. Then there exists an eigenvalue $\lambda^*(\mathbf{X})$
(Perron-Frobenius eigenvalue or PF-eigenvalue) such that
\begin{itemize}

\item[(i)]  $\lambda^*(\mathbf{X})>0 $ and it is real.

\item[(ii)]  there is a positive vector $\mathbf{v}$ such that
$\mathbf{Xv}=\lambda^*(\mathbf{X})\mathbf{v}$.

\item[(iii)]  $\lambda^*(\mathbf{X})>\vert \lambda(\mathbf{X})
\vert $ for any eigenvalue $\lambda(\mathbf{X}) \neq
\lambda^*(\mathbf{X})$.

\item[(iv)]  If $\mathbf{X} \geq \mathbf{Y} \geq \mathbf{0}$, then
$\lambda^*(\mathbf{X}) \geq \vert \lambda(\mathbf{Y}) \vert$ for
any eigenvalue of $\mathbf{Y}$.

\item[(v)]  $\lambda^*(\mathbf{X})$ is a simple root of the
characteristic polynomial of $\mathbf{X}$.
\end{itemize}
\end{theorem}

\begin{lemma}\label{r(u_general)}
The smallest positive root of  $\mathrm{u}_{\Omega}(\gamma)$ is
\begin{equation}
\nonumber
r(\mathrm{u}_{\Omega})=\dfrac{1}{\lambda^*\big(\psi(\mathrm{diag}{(\boldsymbol{\mu})}\mathbf{A},\dfrac{\boldsymbol{\eta}}{\overline{p}_{\Omega}},\Omega)\big)}.
\end{equation}
\end{lemma}

\begin{proof}

\begin{align}
\nonumber\mathrm{u}_{\Omega}(\gamma)
&=\overline{p}_{\Omega}\det{\big(\psi(\mathbf{F},-\dfrac{\gamma\boldsymbol{\eta}}{\overline{p}_{\Omega}},\Omega)\big)}\\
\nonumber &=\overline{p}_{\Omega}\det{\big(\psi(\mathbf{I}-\gamma\mathrm{diag}{(\boldsymbol{\mu})}\mathbf{A},-\dfrac{\gamma\boldsymbol{\eta}}{\overline{p}_{\Omega}},\Omega)\big)}\\
\nonumber &=\overline{p}_{\Omega}\gamma^n\det{\big(\psi(\dfrac{1}{\gamma}\mathbf{I}-\mathrm{diag}{(\boldsymbol{\mu})}\mathbf{A},-\dfrac{\boldsymbol{\eta}}{\overline{p}_{\Omega}},\Omega)\big)}\\
\nonumber
&=\overline{p}_{\Omega}\gamma^n\det{\big(\dfrac{1}{\gamma}\mathbf{I}-\psi(\mathrm{diag}{(\boldsymbol{\mu})}\mathbf{A},\dfrac{\boldsymbol{\eta}}{\overline{p}_{\Omega}},\Omega)\big)}.
\end{align}
Consequently,
$\dfrac{\mathrm{u}_{\Omega}(\gamma)}{\overline{p}_{\Omega}\gamma^n}$
is the reciprocal of the characteristic polynomial of the matrix
$\psi(\mathrm{diag}{(\boldsymbol{\mu})}\mathbf{A},\dfrac{\boldsymbol{\eta}}{\overline{p}_{\Omega}},\Omega)$.
Therefore, the roots of this polynomial are equal to the inverse
of the eigenvalues of
$\psi(\mathrm{diag}{(\boldsymbol{\mu})}\mathbf{A},\dfrac{\boldsymbol{\eta}}{\overline{p}_{\Omega}},\Omega)$.
%Then,
%\begin{equation}
%\nonumber
%\dfrac{1}{\gamma}=\lambda\big(\psi(\mathrm{diag}{(\boldsymbol{\mu})}\mathbf{A},\dfrac{\boldsymbol{\eta}}{\overline{p}_{\Omega}},\Omega)\big).
%\end{equation}
On the other hand, according to Theorem \ref{PF}, since
$\psi(\mathrm{diag}{(\boldsymbol{\mu})}\mathbf{A},\dfrac{\boldsymbol{\eta}}{\overline{p}_{\Omega}},\Omega)$
is a primitive matrix, the PF-eigenvalue of this matrix is real
and positive and has the largest norm among all eigenvalues. Also
it is the simple root of the characteristic polynomial of the
aforementioned matrix. Therefore, the inverse of this eigenvalue
gives the smallest positive simple root of
$\mathrm{u}_{\Omega}(\gamma)$ and the claim is proved.
\end{proof}

%%%%%%%%%%%%%denominator%%%%%%%%%%%%

For the denominator using \eqref{F_definition}, we have
\begin{align}
%\label{denominator}
\nonumber
\mathrm{f}(\gamma)&=\det{(\mathbf{F})}=\det{\big(\mathbf{I}-\gamma\mathrm{diag}{(\boldsymbol{\mu})}\mathbf{A}\big)}\\
         &=\gamma^n\det{\big(\dfrac{1}{\gamma}\mathbf{I}-\mathrm{diag}{(\boldsymbol{\mu})}\mathbf{A}\big)}.
\end{align}
Therefore, $\mathrm{f}(\gamma)$ is the reciprocal of the
characteristic polynomial of
$\mathrm{diag}{(\boldsymbol{\mu})}\mathbf{A}$. On the other hand,
according to Theorem \ref{PF}, the PF-eigenvalue of
$\mathrm{diag}{(\boldsymbol{\mu})}\mathbf{A}$, is real and
positive. It also has the largest magnitude (norm) among the
eigenvalues of the matrix and it is the simple root of the
characteristic polynomial of the associated matrix. Therefore,
$\lambda^*\big(\mathrm{diag}{(\boldsymbol{\mu})}\mathbf{A}\big)$
is the inverse of the smallest positive simple root of
$\mathrm{diag}{(\boldsymbol{\mu})}\mathbf{A}$. Thus,
\begin{equation}\label{f_root}
r(\mathrm{f})=\dfrac{1}{\lambda^*{\big(\mathrm{diag}{(\boldsymbol{\mu})}\mathbf{A}\big)}}.
\end{equation}
On the other hand, according to \eqref{max_u}, $r(\mathrm{f})$ is
also the maximum achievable SINR for the system with unbounded
powers satisfying constraint \eqref{general_constraint1}.
Consequently, using \eqref{min(f,u_general)}, \eqref{f_root} and
Lemma(\ref{r(u_general)}), the maximum achievable SINR to satisfy
all constraints on the power (constraints \eqref{general_constraint1} and
\eqref{general_constraint2}) is
 \begin{align}
\nonumber\gamma^*&=\min{\{ r(\mathrm{f}),{ r(\mathrm{u}^{(i)})\}}}\\
\nonumber
&=\min\{{\dfrac{1}{\lambda^*{\big(\mathrm{diag}{(\boldsymbol{\mu})}\mathbf{A}\big)}},{\dfrac{1}{
\nonumber
\lambda^*\big(\psi(\mathrm{diag}{(\boldsymbol{\mu})}\mathbf{A},\dfrac{\boldsymbol{\eta}}{\overline{p}_i},\{i\})\big)}}}
\}.
\end{align}
Since $\psi
\big(\mathrm{diag}{(\boldsymbol{\mu})}\mathbf{A},\dfrac{\boldsymbol{\eta}}{\overline{p}_i},\Omega\big)
\geq \mathrm{diag}{(\boldsymbol{\mu})}\mathbf{A}$ and both are primitive, using Theorem \ref{PF} we have
\begin{equation}
\nonumber
\lambda^*{\big(\psi(\mathrm{diag}{(\boldsymbol{\mu})}\mathbf{A},\dfrac{\boldsymbol{\eta}}{\overline{p}_i},\{i\})\big)}
\geq
\lambda^*{\big(\mathrm{diag}{(\boldsymbol{\mu})}\mathbf{A}\big)},
\end{equation}
and consequently the maximum achievable $\gamma$ for a system with
constraint on the total power of any subset of the users is
achieved.
%It is easy to see that if the users under the total
%power constraint are not the first $k$ users and randomly
%selected, using a permutation matrix they can be reordered to the
%form with first $k$ users under the total power constraint. Since
%permutation operation doesn't change the value of eigenvalue, at
%the end the users again permuted to their first original ordering.
This discussion leads to the following theorem.

\begin{theorem}\label{general_form}
The maximum achievable $\gamma$ in an interference channel with $n$ links and gain matrix $\mathbf{A}$, where  power vector is subject to the following constraints,
\begin{align}
\nonumber &\mathbf{p}\geq \mathbf{0},\\ \nonumber &\displaystyle\sum_{i\in \Omega}{p_i} \leq \overline{p}_{\Omega}
\end{align}
is equal to
$$
\gamma^*=\dfrac{1}{\lambda^*\big(\psi(\mathrm{diag}{(\boldsymbol{\mu})}\mathbf{A},\dfrac{\boldsymbol{\eta}}{\overline{p}_{\Omega}},\Omega)\big)},
$$

\noindent where $\Omega \subseteq \{1, \ldots ,n
\}$ is an arbitrary subset of the users.
\end{theorem}

When multiple constraints on power exist, it is obvious that the
maximum achievable SINR is the minimum of the maximum achievable
SINR when each of the constraints is applied separately, i.e.,
\begin{equation}\label{multiple_constraints}
\gamma^*=\displaystyle\min_i{\gamma^*_i},
\end{equation}
where $\gamma^*_i$ is the maximum achievable SINR for the
constraint $i$ on power. The following corollary yields the
maximum achievable SINR when the power of individual users and the total power are constrained.

\begin{figure}[bmpt]
\centerline{\psfig{figure=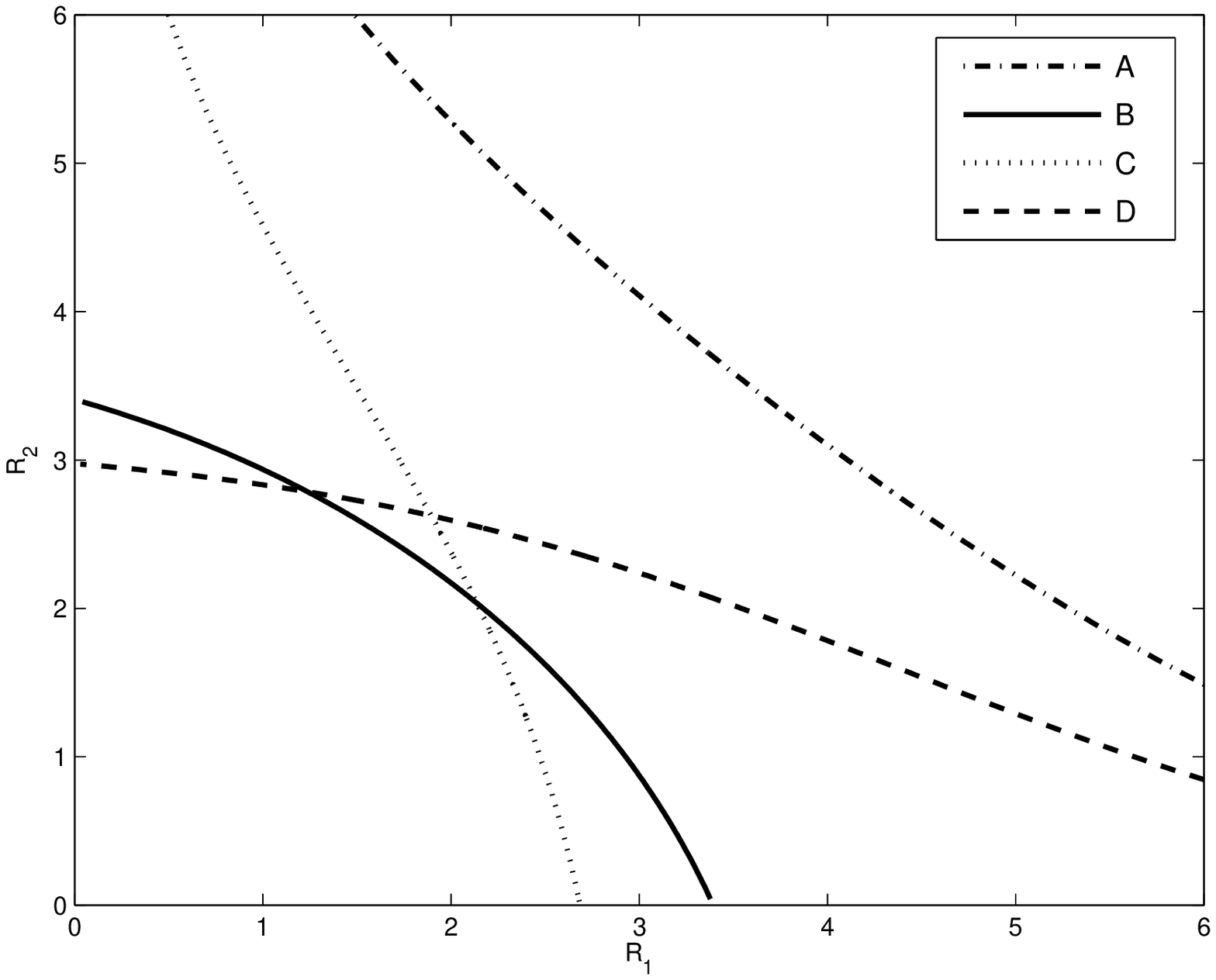,width=3.25 in,height=2.75 in}}
\caption[The rate region for a $2$-user interference channel]{\small{The rate region for a $2$-user interference channel with the following constraints on the power, A: $p_1\geq0$, $p_2\geq0$,    B: $p_1+p_2\geq\bar{p}_t$, $p_1\geq0$, $p_2\geq0$    C: $0\leq p_1 \leq \bar{p}_1$, $p_2\geq 0 $,    D: $0\leq p_2 \leq \bar{p}_2$, $p_1\geq 0 $}}
\label{rr1}
\end{figure}

\begin{figure}[bmpt]
\centerline{\psfig{figure=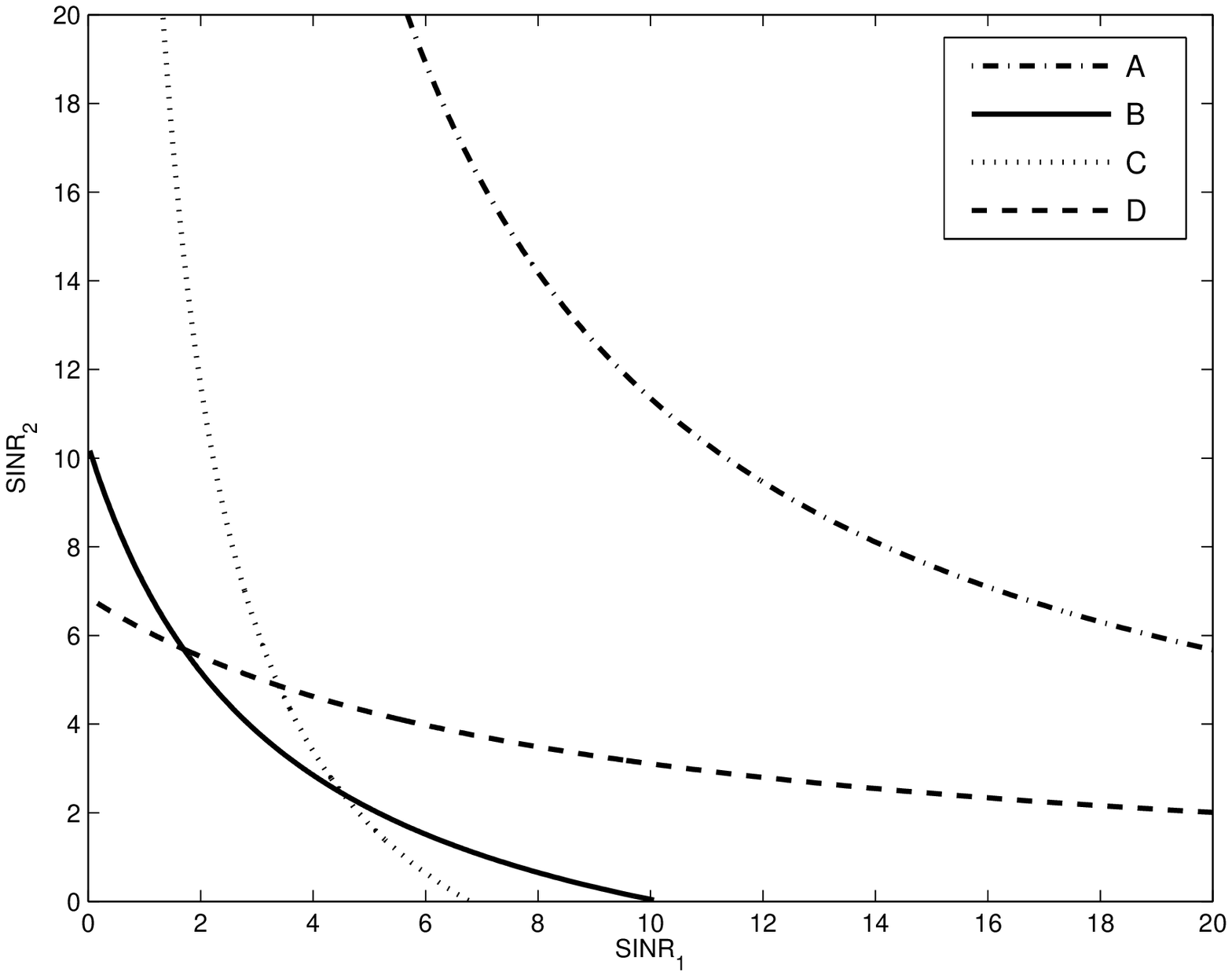,width=3.25 in,height=2.75 in}}
\caption[The SINR region for a $2$-user interference channel]{\small{The rate region for a $2$-user interference channel with the following constraints on the power, A: $p_1\geq0$, $p_2\geq0$,    B: $p_1+p_2\geq\bar{p}_t$, $p_1\geq0$, $p_2\geq0$    C: $0\leq p_1 \leq \bar{p}_1$, $p_2\geq 0 $,    D: $0\leq p_2 \leq \bar{p}_2$, $p_1\geq 0 $}}
\label{sr1}
\end{figure}

\begin{corollary}\label{corol}
The maximum achievable $\gamma$ in \eqref{main_problem}, where
power vector is subject to the following constraints,
\begin{align}
\nonumber &\mathbf{p}\geq \mathbf{0}, \\ \nonumber & \mathbf{p} \leq \mathbf{\overline{p}}, \\ \nonumber &  \displaystyle\sum_{i=1}^{n}{p_i} \leq \overline{p}_t
\end{align}
is equal to $\nonumber\gamma^*=$
\begin{align}
\min \{&
\dfrac{1}{\lambda^*\big(\psi(\mathrm{diag}{(\boldsymbol{\mu})}\mathbf{A},
\dfrac{\boldsymbol{\eta}}{\overline{p}_t},\{1,\ldots,n\})\big)},\\
\nonumber
&\dfrac{1}{\lambda^*\big(\psi(\mathrm{diag}{(\boldsymbol{\mu})}\mathbf{A},
\dfrac{\boldsymbol{\eta}}{\overline{p}_1},\{1\})\big)}, \\
&\nonumber\dfrac{1}{\lambda^*\big(\psi(\mathrm{diag}{(\boldsymbol{\mu})}\mathbf{A},\dfrac{\boldsymbol{\eta}}{\overline{p}_2},\{2\})\big)},\\
\nonumber &\ldots
,\dfrac{1}{\lambda^*\big(\psi(\mathrm{diag}{(\boldsymbol{\mu})}\mathbf{A},\dfrac{\boldsymbol{\eta}}{\overline{p}_n},\{n\})\big)}
\}.
\end{align}
\end{corollary}
The boundary of the SINR region in any direction can be obtained
by choosing $\boldsymbol{\mu}$, accordingly. Due to the explicit relationship between the SINR and the rate in Gaussian channels, obtaining the SINR region in these channels amounts to the rate region characterization. As an example, Fig.
\ref{rr1} and \ref{sr1}, respectively, depict the rate region and SINR region of a system with the gain matrix
$\mathbf{G}$ as
\begin{equation}
\nonumber
\mathbf{G}=\left[
\begin{array}{ll}
0.6791 & 0.0999\\
0.0411 & 0.6864
\end{array}
\right],
\end{equation}
% \kappa=10^-1
while the power of individual users and the total power are upper-bounded as $\overline{p}_{1}=0.8,\quad \overline{p}_{2}=1, \quad
\overline{p}_t=1.4$, and $\sigma_1^2=\sigma_2^2=10^{-1}$.

\begin{figure}[bmpt]
\centerline{\psfig{figure=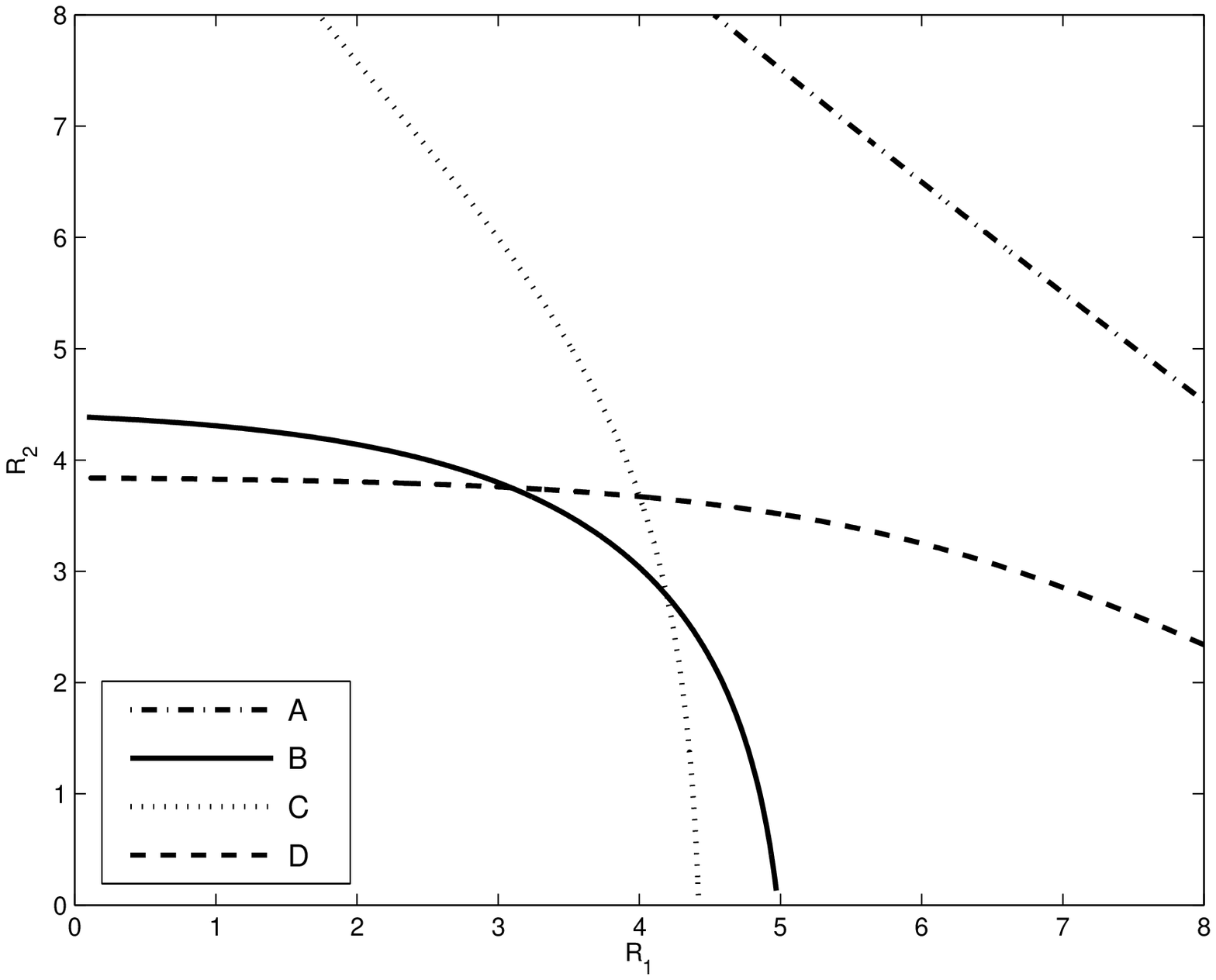,width=3.25 in,height=2.75 in}}
\caption[The rate region for a $2$-user interference channel with weak cross links]{\small{The rate region for a $2$-user interference channel with the following constraints on the power, A: $p_1\geq0$, $p_2\geq0$,    B: $p_1+p_2\geq\bar{p}_t$, $p_1\geq0$, $p_2\geq0$    C: $0\leq p_1 \leq \bar{p}_1$, $p_2\geq 0 $,    D: $0\leq p_2 \leq \bar{p}_2$, $p_1\geq 0 $}}
\label{rr2}
\end{figure}

\begin{figure}[bmpt]
\centerline{\psfig{figure=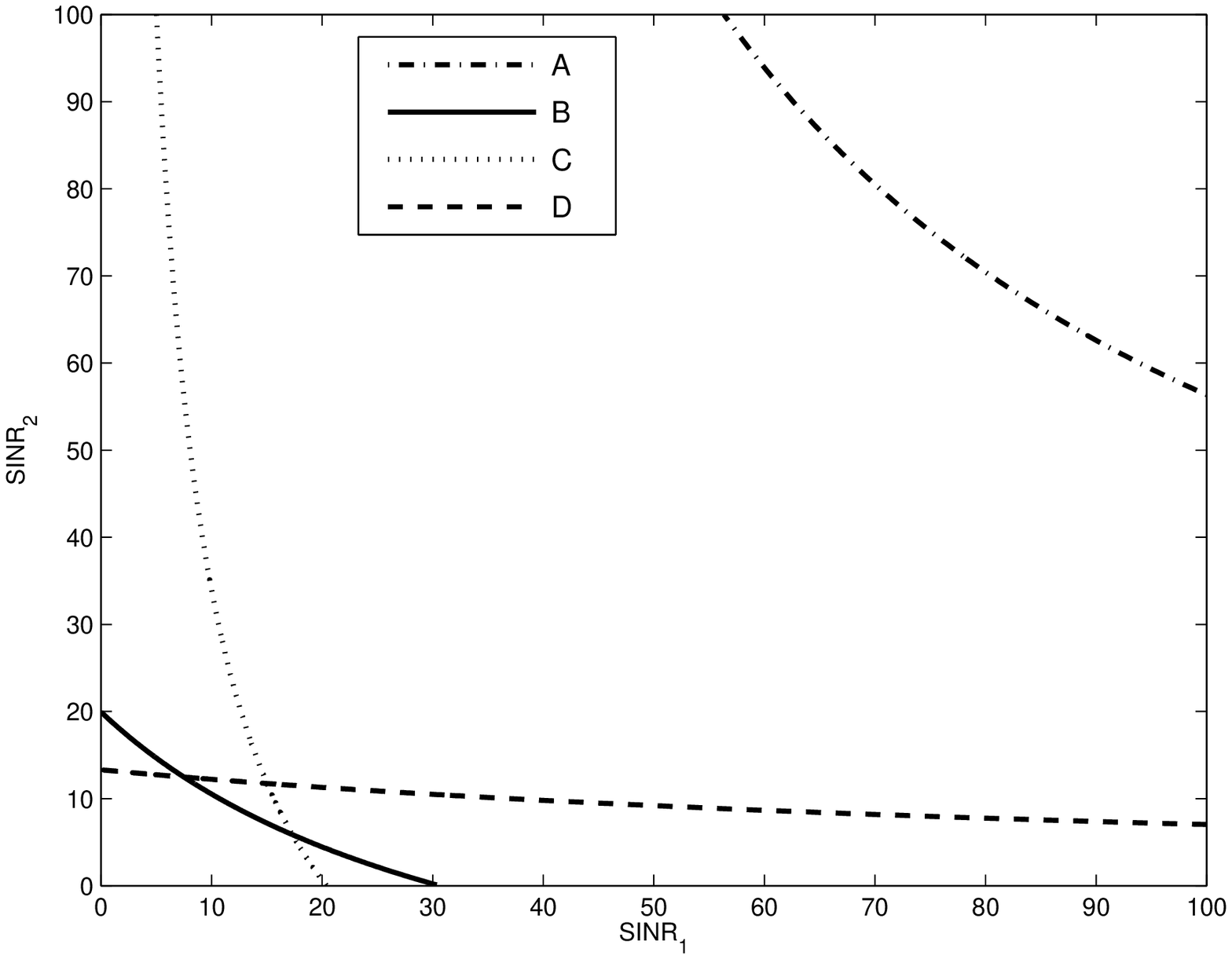,width=3.25 in,height=2.75 in}}
\caption[The SINR region for a $2$-user interference channel with weak cross links]{\small{The rate region for a $2$-user interference channel with the following constraints on the power, A: $p_1\geq0$, $p_2\geq0$,    B: $p_1+p_2\geq\bar{p}_t$, $p_1\geq0$, $p_2\geq0$    C: $0\leq p_1 \leq \bar{p}_1$, $p_2\geq 0 $,    D: $0\leq p_2 \leq \bar{p}_2$, $p_1\geq 0 $ }}
\label{sr2}
\end{figure}
The rate region is simply the intersection of all the rate regions
resulted from applying each constraint separately. As shown in
Fig. \ref{rr1} and Fig. \ref{sr1}, the boundary of SINR and rate
region, when there is no upper-bound on powers is always above
other boundaries. It is because of the fact that the maximum
achievable SINR for the unbounded-power system is the inverse of
PF-eigenvalue of $\mathrm{diag}{(\boldsymbol{\mu})}\mathbf{A}$;
while the maximum achievable SINR when the power is bounded, is
the inverse of PF-eigenvalue of a matrix which is definitely
greater than $\mathrm{diag}{(\boldsymbol{\mu})}\mathbf{A}$.
Therefore, based on Theorem \ref{PF} the unbounded SINR boundary
would be above the bounded-power systems. Thus, this boundary
doesn't have any direct role in forming the main boundary. An interesting
observation is that if the $\overline{p}_i$'s or $\overline{p}_t$
are increased the boundaries of bounded-power systems tend to the
unbounded-power system boundary; the extreme case is when the
maximum power goes to infinity which means the power is unbounded,
then the matrices whose inverse of PF-eigenvalue form the
boundaries become equal and these boundaries touch each other.

 As another observation, the rate and SINR regions for a 2-user channel with weaker cross
links are shown in Fig. \ref{rr2} and \ref{sr2}. The gain matrix
in this system is assumed to be
\begin{equation} \mathbf{G}=\left[
\begin{array}{ll}
2.0430  &  0.0359 \\
0.0134  &  1.3313
\end{array}
\right],
\end{equation}
while the power of individual users and the total power are upper-bounded as $\overline{p}_{1}=1,\quad \overline{p}_{2}=1, \quad
\overline{p}_t=1.5$, and $\sigma_1^2=\sigma_2^2=10^{-1}$.
The extreme point of this situation is when the links have no
interference on each other and therefore the maximum SINR for each
user considering the individual constraints would be
$SINR_i=\dfrac{\overline{p}_ig(i,i)}{\sigma^2}$ for each user. We
can see in Fig. \ref{rr2} and Fig. \ref{sr2} that these boundaries
are more straight than the ones in  Fig. \ref{rr1} and Fig.
\ref{sr1} which confirms our conjecture.

\section{Time-Varying Channel}

So far, we have assumed that the channel gains are fixed with time.
However, in practice, channel gains vary with time due to the users' movement or changing the environment
conditions.

%In this case, instead of instantaneous power the
%average power is constrained.

In this section, we consider an interference channel with $n$ co-channel links
whose channel gain matrix is randomly selected from a finite set  $\{\mathbf{G}_1,\ldots, \mathbf{G}_l \}$
with probability $\rho_1,\ldots,\rho_l$, respectively. The matrix $\mathbf{A}_i$ denotes the normalized gain matrix in the state $i$, $i\in\{1,\ldots,l\}$.
The objective is to find the maximum $\gamma$ which is achievable by all users in all channel states, while the average power of the users are constrained, i.e.,
\begin{align}
 %\label{main_problem_time}
\nonumber&\max {\gamma} \\
%\label{main_constraint_time}
\nonumber \mathrm{s.t.} \quad &\gamma_{j+(i-1)n} \geq \mu_j \gamma, \quad \forall j \in \Omega,i\in \{1,\ldots,l\} \\
\label{general_constraint1_time}
 &p_{j+(i-1)n}\geq 0,\quad \forall j \in \Omega,i\in \{1,\ldots,l\} \\
\label{general_constraint2_time}
&\displaystyle E[\sum_{j\in \Omega}{p_{j+(i-1)n}}] \leq
\overline{p}_{\Omega},
\end{align}
where $\gamma_{j+(i-1)n}$ and $p_{j+(i-1)n}$ are the SINR and the power of transmitter $j$ respectively, when the channel gain matrix is $\mathbf{G}_i$.
We define an expanded system including $ln$ users with
block diagonal matrices $\mathbf{G}$ and $\mathbf{A}$ as the channel gain matrix
and the normalized gain matrix, respectively. In the matrices $\mathbf{G}$ and $\mathbf{A}$, the $i^{th}$ matrix on the diagonal is $\mathbf{G}_i$ and $\mathbf{A}_i$, respectively. It is clear that block diagonal
format of these matrices indicate that there is no interference
between the links associated with different states. Like the
previous discussions the requirements on these links form a system
of linear equations with the following formulation in a matrix form,

\begin{equation}
(\dfrac{1}{\gamma}\mathbf{I}_{ln \times
ln}-\mathrm{diag}{(\mathbf{1}_{l \times
1}\otimes\boldsymbol{\mu})}\mathbf{A})\mathbf{p}=\boldsymbol{\eta},
\end{equation}
where $$\eta_{j+(i-1)n}=\dfrac{\mu_j\sigma^2_j}{g_{j+(i-1)n,j+(i-1)n}}, j \in \Omega, i \in \{1,\ldots,l\}.$$
According to \eqref{F_definition}, we define $\mathbf{F}$ as
\begin{equation}
\nonumber \mathbf{F}=\dfrac{1}{\gamma}\mathbf{I}_{ln \times
ln}-\mathrm{diag}{(\mathbf{1}_{l \times
1}\otimes\boldsymbol{\mu})}\mathbf{A}.
\end{equation}
Then, we have
\begin{equation}
\nonumber \mathbf{F}\mathbf{p}=\gamma\boldsymbol{\eta}.
\end{equation}
Using Cramer's rule, we will have
\begin{equation}
\nonumber p_{j+(i-1)n}=\dfrac{\det(\mathbf{H}^{(j+(i-1)n)})}{\det(\mathbf{F})},
\end{equation}
where $\mathbf{H}^{(j+(i-1)n)}$ according to \eqref{H_definition} is the matrix $\mathbf{F}$ whose column $j+(i-1)n$ is substituted by $\gamma\boldsymbol{\eta}$.
The average of the total power of the users in $\Omega$ is equivalent to
\begin{align}
\nonumber \displaystyle E_i \sum_{j\in \Omega}{p_{j+(i-1)n}}&=\displaystyle\sum_{i=1}^l{\rho_i\displaystyle\sum_{j\in \Omega}{ p_{j+(i-1)n}}}\\
\nonumber &=\displaystyle\sum_{i=1}^l{\rho_i \displaystyle\sum_{j\in \Omega}{\dfrac{\det(\mathbf{H}^{(j+(i-1)n)})}{\det(\mathbf{F})}}}\\
\label{long_inequality} &=\dfrac{1}{\det(\mathbf{F})}\displaystyle\sum_{i=1}^{l}{\rho_i\displaystyle\sum_{j\in
\Omega}{ \det(\mathbf{H}^{(j+(i-1)n)})}} .
\end{align}
Based on \eqref{long_inequality}, we define
\begin{equation}\label{u_time}
\nonumber\mathrm{u}_{\Omega}(\gamma)=\overline{p}_{\Omega}\det(\mathbf{F})-\displaystyle\sum_{i=1}^{l}{\rho_i\displaystyle\sum_{j\in
\Omega}{ \det(\mathbf{H}^{(j+(i-1)n)})}},
\end{equation}
and
\begin{equation}\label{f_time}
\nonumber \mathrm{f}(\gamma)=\det(\mathbf{F}).
\end{equation}
Therefore, the constraint in \eqref{general_constraint2_time} is equivalent to
\begin{align}
\nonumber \dfrac{\mathrm{u}_{\Omega}(\gamma)}{\mathrm{f}(\gamma)} \geq 0 .
\end{align}
Like before, it is easy to show that the maximum achievable SINR
satisfying constraints \eqref{general_constraint1_time} and
\eqref{general_constraint2_time} is
\begin{equation}\label{gamma_min_time}
\gamma^*=\min{\{r(\mathrm{f}),r(\mathrm{u}_{\Omega})\}}.
\end{equation}
To simplify $\mathrm{u}_{\Omega}(\gamma)$, we have
\begin{align}
\nonumber \mathrm{u}_{\Omega}(\gamma)
&=\overline{p}_{\Omega}\det(\mathbf{F})-\displaystyle\sum_{i=1}^{l}{\rho_i\displaystyle\sum_{j\in
\Omega}{
\det(\mathbf{H}^{(j+(i-1)n)})}} \\
\nonumber &=\overline{p}_{\Omega}(\det(\mathbf{F})-\displaystyle\sum_{i=1}^{l}{\displaystyle\sum_{j\in
\Omega}{ \det(\mathbf{\hat{H}}^{(j+(i-1)n)})}}),
\end{align}
where $\mathbf{\hat{H}}^{(j+(i-1)n)}$ is
$\mathbf{{H}}^{(j+(i-1)n)}$ whose column $j+(i-1)n$ is multiplied
by $\dfrac{\rho_i}{\overline{p}_{\Omega}}$. Using the same
procedure as before, we obtain
\begin{equation}
\nonumber \mathrm{u}_{\Omega}(\gamma)=\overline{p}_{\Omega}\det{(\mathbf{F}-\mathbf{D})},
\end{equation}
where
\begin{equation}
\nonumber \mathbf{D}=\displaystyle\sum_{i=1}^{l}{\psi(\mathbf{0}_{ln \times
ln},\dfrac{\rho_i\gamma\boldsymbol{\eta}}{\overline{p}_{\Omega}},\{j+(i-1)n:j
\in \Omega \})}
\end{equation}
It is easy to see that
\begin{equation}
 \nonumber r(\mathrm{u}_{\Omega})=\dfrac{1}{\lambda^*\big(\mathrm{diag}{(\mathbf{1}_{l \times 1} \otimes
 \boldsymbol{\mu})}\mathbf{A}+\displaystyle\sum_{i=1}^{l}{\psi(\mathbf{0}_{ln \times ln},\dfrac{\rho_i\boldsymbol{\eta}}{\overline{p}_{\Omega}},\{j+(i-1)n:j
\in \Omega \})}\big)}.
\end{equation}
and
\begin{equation}
 \nonumber r(\mathrm{f})=\dfrac{1}{\lambda^*(\mathrm{diag}{(\mathbf{1}_{l \times
1}\otimes\boldsymbol{\mu})}\mathbf{A})}.
\end{equation}
Therefore, using Theorem \ref{PF} and equation \eqref{gamma_min_time}, we will have
the following theorem.
\begin{theorem}\label{general_form_time_varying}
The maximum achievable $\gamma$ in a time-varying interference channel with $n$ links and probability vector $\boldsymbol{\rho}_{l\times1}$,
 with the following constraints on power,
\begin{align}
\nonumber
&p_{j,i}\geq 0, \forall j \in \Omega, i \in \{1,\ldots,l\}, \quad \\
\nonumber &\displaystyle E[\sum_{j\in \Omega}{p_{j,i}}] \leq
\overline{p}_{\Omega}
\end{align}
is equal to
\begin{align}
\nonumber
\gamma^*=\dfrac{1}{\lambda^*\big(\mathrm{diag}{(\mathbf{1}_{l \times 1} \otimes
 \boldsymbol{\mu})}\mathbf{A}+\displaystyle\sum_{i=1}^{l}{\psi(\mathbf{0}_{ln \times ln},\dfrac{\rho_i\boldsymbol{\eta}}{\overline{p}_{\Omega}},\{j+(i-1)n:j
\in \Omega \})}\big)}.
\end{align}
\end{theorem}
Apparently, if there are multiple constraints on the power, the maximum achievable SINR $\gamma^*$ is computed by
$$
\gamma^*=\displaystyle\min_i{\gamma^*_i},
$$
where $\gamma^*_i$ is the maximum achievable SINR obtained by Theorem \ref{general_form_time_varying} while only the constraint $i$ is considered for the system.

%Although the size of the matrix whose inverse of eigenvalue yields
%the maximum achievable SINR for the system grows with $n$ and $l$,
%this is a closed-form solution for the rate region of such a
%system which is very valuable; while applying it to the small
%systems with small enough states is feasible. In addition, this
%matrix has special structure, some approximation methods which
%reduce the computational complexity of such a method may exist.

\section{Conclusion}
\comment{ Interference channels and their application have been
emerging in the new wireless communication networks technology
ubiquitously. In spite of the benefits of high capacity and
coverage, the interference on the co-channel signals cause a
deterioration in the performance of the system. Although many
interference reducing techniques such as sectorization, smart
antennas, interference averaging, multiuser detection, and
interference precancelation try to mitigate this problem, the large
complexity of these systems doesn't allow it to be practical.
Moreover, many of these methods can not remove the interference
completely and still the system's performance is compromised.
Consequently, the existence of such interference limits the QoS to a
maximum value. }
In this paper, we have obtained a closed-form solution for the maximum achievable SINR
in an interference channel, utilizing the Perron-Frobenious
theorem, when there is a total power constraint on any subset of
the users. This result leads to characterizing the boundary of the
rate region with multiple constraints on the power. In addition, we considered a time-varying interference channel where
the average of total power of an arbitrary subset of the transmitters is
subject to an upper-bound. A closed-form expression for the rate-region of such a channel is obtained and extended to the systems with multiple power constraints.

\bibliographystyle{IEEE}

\end{document}